# BAYESBALL: A BAYESIAN HIERARCHICAL MODEL FOR EVALUATING FIELDING IN MAJOR LEAGUE BASEBALL

By Shane T. Jensen, Kenneth E. Shirley and Abraham J. Wyner

*University of Pennsylvania*

The use of statistical modeling in baseball has received substantial attention recently in both the media and academic community. We focus on a relatively under-explored topic: the use of statistical models for the analysis of fielding based on high-resolution data consisting of on-field location of batted balls. We combine spatial modeling with a hierarchical Bayesian structure in order to evaluate the performance of individual fielders while sharing information between fielders at each position. We present results across four seasons of MLB data (2002–2005) and compare our approach to other fielding evaluation procedures.

**1. Introduction.** Many aspects of major league baseball are relatively easy to evaluate because of the mostly discrete nature of the game: there are a relatively small number of possible outcomes for each hitting or pitching event. In addition, it is easy to determine which player is responsible for these outcomes. Complicating and confounding factors exist—like ball parks and league—but these differences are either small or averaged out over the course of a season.

A player's fielding ability is more difficult to evaluate, because fielding is a nondiscrete aspect of the game, with players fielding balls-in-play (BIPs) across the continuous playing surface. Each ball-in-play is either successfully fielded by a defensive player, leading to an out (or multiple outs) on the play, or the ball-in-play is not successfully fielded, resulting in a hit. An inherently complicated aspect of fielding analysis is assessing the blame for an unsuccessful fielding play. Specific unsuccessful fielding plays can be deemed to be an "error" by the official scorer at each game. These assigned errors are easy to tabulate and can be used as a rudimentary measure for comparing players. However, errors are a subjective measure [Kalist and Spurr (2006)] that only tell part of the story. Additionally, errors are only reserved for plays where a ball-in-play is obviously mishandled, with no corresponding measure for rewarding players for a particularly well-handled fielding play. Most analysts agree that a more objective measure of fielding ability is the range of the fielder, though this quality is hard to measure. If a batted ball sneaks through the left side of the infield, for example, it is very difficult to know if a faster or better positioned shortstop could have reasonably made the play. Confounding factors such as the speed and trajectory of the batted ball and the quality and range of adjacent fielders abound.

Furthermore, because of the large and continuous playing surface, the evaluation of fielding in major league baseball presents a greater modeling challenge than the evaluation of offensive contributions. Previous approaches have addressed this problem by avoiding continuous models and instead discretizing the playing surface. The Ultimate Zone Rating (UZR) is based on a division of the playing field into 64 large zones, with fielders evaluated by tabulating their successful plays within each zone [Lichtman (2003)]. The Probabilistic Model of Range (PMR) divides the field into 18 pie slices (every 5 degrees) on either side of second base, with fielders evaluated by tabulating their successful plays within each slice [Pinto (2006)]. Another similar method is the recently published Plus-Minus system [Dewan (2006)]. The weakness of these methods is that each zone or slice is quite large, which limits the extent to which differences between fielders are detectable, since every ball hit into a zone is treated equally.

Our methodology addresses the continuous playing surface by modeling the success of a fielder on a given BIP as a function of the location of that BIP, where location is measured as a *continuous* variable. We fit a hierarchical Bayesian model to evaluate the success of each individual fielder, while sharing information between fielders at the same position. Hierarchical Bayesian models have also recently been used by Reich et al. (2006) to estimate the spatial distribution of basketball shot chart data. Our ultimate goal is to produce an evaluation by estimating the number of runs that a given fielder saves or costs his team during the season compared to the average fielder at his position. Since this quantity is not directly observed, it cannot be used as the outcome variable in a statistical model. Therefore, our evaluation requires two steps. First, we model the binary variable of whether a player successfully fields a given BIP (an outcome we can observe) as a function of the BIP location. Then, we integrate over the estimated distribution of BIP locations and multiply by the estimated consequence of a successful or unsuccessful play, measured in runs, to arrive at our final estimate of the number of runs saved or cost by a given fielder in a season.

We present our Bayesian hierarchical model implemented on high-resolution data in Section 2. In Section 3 we illustrate our method using one particular position and BIP type as an example. In Section 4 we describe the calculations we make to convert the parameter estimates from the Bayesian hierarchical model to an estimate of the runs saved or cost. In Section 5 we present our integrated results, and we compare our results to those from a representative previous method, UZR, in Section 6. We conclude with a discussion in Section 7.

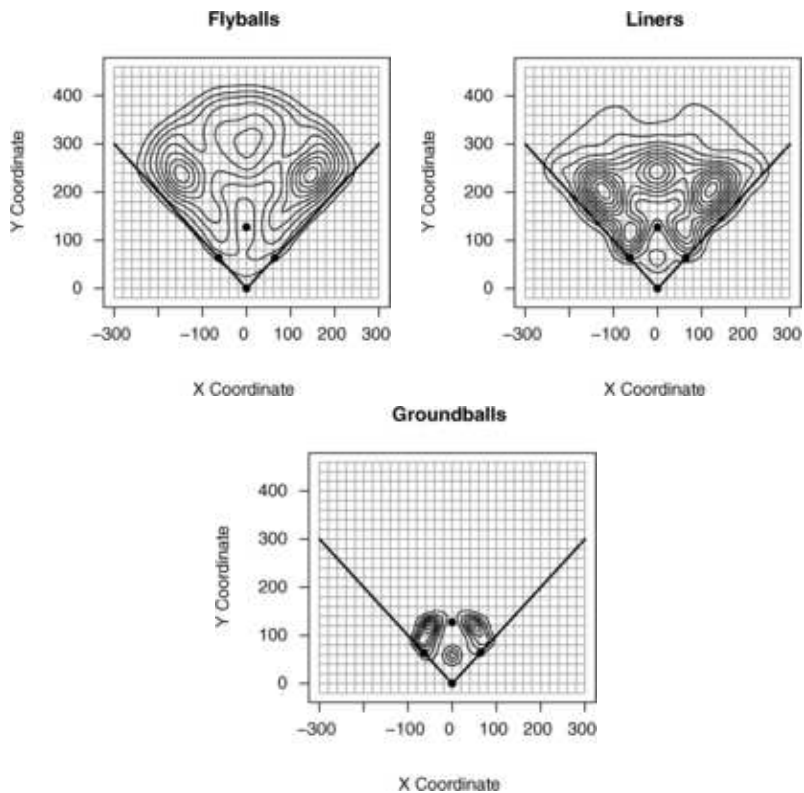

FIG. 1. *Contour plots of estimated 2-dimensional densities of the 3 BIP types, using all data from 2002–2005. Note that the origin is located at home plate, and the four bases are drawn into the plots as black dots, where the diagonal lines are the left and right foul lines. The outfield fence is not drawn into the plot, because the data come from multiple ballparks, each with its outfield fence in a different place. The units of measurement for both axes are feet.*

## 2. Bayesian hierarchical model for individual players.

2.1. *The data.* Our fielding evaluation is based upon high-resolution data collected by Baseball Info Solutions [BIS (2007)]. Every ball put into play in a major league baseball game is mapped to an $(x, y)$ coordinate on the playing field, up to a resolution of approximately $4 \times 4$ feet. Our research team collected samples from several companies that provide high-resolution data and after watching replays of several games, we decided to use the BIS data since it appeared to be the most accurate. We have four seasons of data (2002–2005), with around 120,000 balls-in-play (BIP) per year. These BIPs are classified into three distinct types: flyballs (33% of BIP), liners (25% of BIP) and grounders (42% of BIP). The flyballs category also includes infield and outfield pop-ups. Figure 1 displays the estimated 2-dimensional density of each of the three BIP types, plotted on the 2-dimensional playing surface. The areas of the field with the highest density of balls-in-play are indicated by the contour lines which are in closest proximity to each other. Not surprisingly, the high-density BIP areas are quite different between the three BIP types. For flyballs and liners, the location of each BIP is the $(x, y)$-coordinate where the ball was either caught (if it was caught) or where the ball landed (if it was not caught). For grounders, the $(x, y)$-location of the BIP is set to the location where the grounder was fielded, either by an infielder or an outfielder (if the ball made it through the infield for a hit).

Table 1
*Summary of models*

| BIP-type | Flyballs | Liners | Grounders |
|---|---|---|---|
| Position | 1B | 1B | 1B |
|  | 2B | 2B | 2B |
|  | 3B | 3B | 3B |
|  | SS | SS | SS |
|  | LF | LF |  |
|  | CF | CF |  |
|  | RF | RF |  |

2.2. *Overview of our models.* The first goal of our analysis is to probabilistically model the binary outcome of whether a fielder made a "successful play" on a ball batted into fair territory. We fit a separate model for each combination of year (2002–2005), BIP type (flyball, liner, grounder) and position. Table 1 contains a listing of the models we fit classified by position and BIP type. Pitchers and catchers were excluded due to a lack of data. Also note that fly balls and liners are modeled for all seven remaining positions, whereas grounders are only modeled for the infield positions. This gives us eighteen models to be fit within each of the four years, giving us $18 \times 4 = 72$ total model fits. The inputs available for modeling include the identity of the fielder playing the given position, the location of the batted ball, and the approximate velocity of the batted ball, measured as an ordinal variable with three levels (the velocity variable is estimated by human observation of video, not using any machinery). For flyballs and liners, a successful play is defined to be a play in which the fielder catches the ball in the air before it hits the ground. For grounders, a successful play is defined to be a play in which the fielder fields the grounder and records at least one out on the play. Grounders and Flyballs/Liner BIPs are fundamentally different in the way their location data is recorded, as outlined below, which affects our modeling approach.

1. *Flyballs and liners*: For flyballs and liners, the $(x, y)$-location of the BIP is set to the location where the ball was either caught (if it was caught) or where the ball landed (if it was not caught). We model the probability of a catch as a function of the distance a player had to travel to reach the BIP location, the direction he had to travel (forward or backward) and the velocity of the BIP. Our flyball/liner distances must incorporate two dimensions since a fielder travels across a two-dimensional plane (the playing field) to catch the BIP.
2. *Grounders*: For grounders, the $(x, y)$-location of the BIP is set to the location where the grounder was fielded, either by an infielder or an outfielder (if the ball made it through the infield for a hit). As we did with flyballs/liners, we model the probability of an infielder successfully fielding a grounder as a function of the distance, direction and velocity of the grounder. For grounders, however, distance is measured as the angle, in degrees, between the trajectory of the groundball from home plate and the (imaginary) line drawn between the infielder's starting location and home plate, with direction being factored in by allowing different probabilities for fielders moving the same number of degrees to the left or the right. The grounder distance only must incorporate one dimension since the infielder travels along a one-dimensional path (arc) in order to field a grounder BIP.

Figure 2 gives a graphical representation of the difference in our approach between grounders and flys/liners. It is worth noting, however, that the distance (for flyballs/liners) or angle (for grounders) that a fielder must travel in order to reach a BIP is actually an estimated value, since the actual starting location of the fielder for any particular play is not included in the data. Instead, the starting location for each position is estimated as the location in the field where each position has the highest overall proportion of successful plays. The distance/angle traveled for each BIP is then calculated relative to this estimated starting position for each player.

2.3. *Model for flyballs/liners using a two-dimensional spatial representation.* We present our model below in the context of flyballs (which also include infield pop-ups), but the same methodology is used for liners as well. For a particular fielder $i$, we denote the number of BIPs hit while that player was playing defense $n_i$. The outcome of each play is either a success

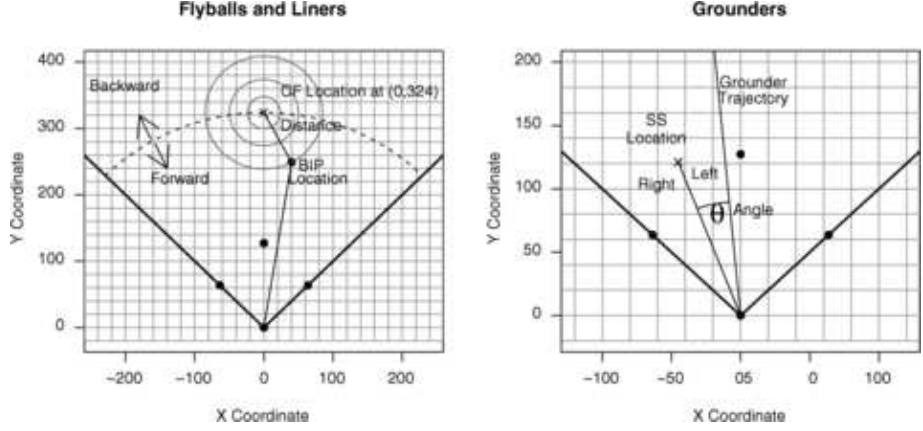

FIG. 2. *Two-dimensional representation for flyballs and liners versus one-dimensional representation for grounders.*

or failure:

$$S_{ij} = \begin{cases} 1, & \text{if the } j\text{th flyball hit to the } i\text{th player is caught,} \\ 0, & \text{if the } j\text{th flyball hit to the } i\text{th player is not caught.} \end{cases}$$

These observed successes and failures are modeled as Bernoulli realizations from an underlying event-specific probability:

(1) $$S_{ij} \sim \text{Bernoulli}(p_{ij}).$$

As mentioned above, the available covariates are the $(x, y)$ location and the velocity $V_{ij}$ of the BIP. Although the velocity is an ordinal variable $V_{ij} = \{1, 2, 3\}$, we treat velocity as a continuous variable in our model in order to reduce the number of coefficients included. The Bernoulli probabilities $p_{ij}$ are modeled as a function of distance $D_{ij}$ traveled to the BIP, velocity $V_{ij}$ and an indicator for the direction $F_{ij}$ the fielder has to move toward the BIP ($F_{ij} = 1$ for moving forward, $F_{ij} = 0$ for moving backward):

(2) $$\begin{aligned} p_{ij} &= \Phi(\beta_{i0} + \beta_{i1} D_{ij} + \beta_{i2} D_{ij} F_{ij} + \beta_{i3} D_{ij} V_{ij} + \beta_{i4} D_{ij} V_{ij} F_{ij}) \\ &= \Phi(\mathbf{X}_{ij} \cdot \boldsymbol{\beta}_i), \end{aligned}$$

where $\Phi(\cdot)$ is the cumulative distribution function for the Normal distribution and $\mathbf{X}_{ij}$ is a vector of the covariate terms in equation (2). Note that the covariates $D_{ij}$ and $F_{ij}$ are themselves functions of the $(x, y)$ coordinates for that particular BIP. This model is recognizable as a probit regression model with interactions between covariates that allow for different probabilities for moving the same distance in the forward direction versus the backward direction. We can give natural interpretations to the parameters of this fly/liner probit model. The $\beta_{i0}$ parameter controls the probability of a catch on a fly/liner hit directly at a fielder ($D_{ij} = 0$). The $\beta_{i1}$ and $\beta_{i2}$ parameters control the range of the fielder, moving either backward ($\beta_{i1}$) or forward ($\beta_{i2}$) toward a fly/liner. The parameters $\beta_{i3}$ and $\beta_{i4}$ adjust the probability of success as a function of velocity.

2.4. *Model for grounders using a one-dimensional spatial representation.* The outcome of each grounder BIP is either a success or failure:

$$S_{ij} = \begin{cases} 1, & \text{if the } j\text{th grounder hit to the } i\text{th player is fielded successfully,} \\ 0, & \text{if the } j\text{th grounder hit to the } i\text{th player is not fielded successfully.} \end{cases}$$

Grounders have a similar observed data level to their model,

$$(3) \qquad S_{ij} \sim \text{Bernoulli}(p_{ij}),$$

except that the underlying probabilities $p_{ij}$ are modeled as a function of angle $\theta_{ij}$ between the fielder and the BIP location, the velocity $V_{ij}$ of the BIP, and an indicator for the direction $L_{ij}$ the fielder has to move toward the BIP ($L_{ij} = 1$ for moving to the left, $L_{ij} = 0$ for moving to the right):

$$(4) \qquad \begin{aligned} p_{ij} &= \Phi(\beta_{i0} + \beta_{i1}\theta_{ij} + \beta_{i2}\theta_{ij}L_{ij} + \beta_{i3}\theta_{ij}V_{ij} + \beta_{i4}\theta_{ij}V_{ij}L_{ij}) \\ &= \Phi(\mathbf{X}_{ij} \cdot \boldsymbol{\beta}_i). \end{aligned}$$

Again $\Phi(\cdot)$ represents the cumulative distribution function for the Normal distribution and $\mathbf{X}_{ij}$ is a vector of the covariate terms in equation (4). We can also give natural interpretations of the parameters in this grounder probit model. The $\beta_{i0}$ parameter controls the probability of a catch on a grounder hit directly at the fielder ($D_{ij} = 0$). The $\beta_{i1}$ and $\beta_{i2}$ parameters control the range of the fielder, moving either to the right ($\beta_{i1}$) or to the left ($\beta_{i2}$) toward a grounder. The parameters $\beta_{i3}$ and $\beta_{i4}$ adjust the probability of success as a function of velocity.

2.5. *Sharing information between players.* We can calculate parameter estimates $\boldsymbol{\beta}_i$ for each player $i$ separately using standard probit regression software. However, we will see in Section 3.2 below that these parameter estimates $\boldsymbol{\beta}_i$ can be highly variable for players with small sample sizes (i.e., those players who faced a small number of BIPs in a given year). This problem can be addressed by using a hierarchical model where each set of player-specific coefficients $\boldsymbol{\beta}_i$ are modeled as sharing a common prior distribution. This hierarchical structure allows for information to be shared between all players at a position, which is especially important for players with smaller numbers of opportunities. Specifically, we model each player-specific coefficient as a draw from a common distribution shared by all players at a position:

$$(5) \qquad \boldsymbol{\beta}_i \sim \text{Normal}(\boldsymbol{\mu}, \Sigma),$$

where $\boldsymbol{\mu}$ is the $5 \times 1$ vector of means and $\Sigma$ is the $5 \times 5$ prior covariance matrix shared across all players. We assume a priori independence of the components of $\boldsymbol{\beta}_i$, so that $\Sigma$ has off-diagonal elements of zero, and diagonal elements of $\sigma_k^2$ ($k = 0, \ldots, 4$). Although the components of $\boldsymbol{\beta}_i$ are assumed to be independent a priori, there will be posterior dependence between these components induced by the data. The functional form of this posterior dependence is given in our supplementary materials section on model implementation [Jensen, Shirley and Wyner (2009)]. Finally, we must also specify a prior distribution for the shared player parameters $(\mu_k, \sigma_k : k = 0, \ldots, 4)$, which we choose to be noninformative following the recommendation of Gelman (2006),

$$(6) \qquad p(\mu_k, \sigma_k) \propto 1, \qquad k = 0, \ldots, 4.$$

We also explored the use of alternative prior specifications, including a proper inverse-Gamma prior distribution for $\sigma_k^2$: $(\sigma_k^2)^{-1} \sim \text{Gamma}(a, b)$, where $a$ and $b$ are small values ($a = b = 0.0001$). We observed very little difference in our posterior estimates using this alternative prior distribution.

For each position and BIP type, our full set of unknown parameters are $\boldsymbol{\beta}$, the $N \times 5$ matrix containing the coefficients of each player at a particular position ($N$ = number of players at that position), as well as $\boldsymbol{\mu}$, the $5 \times 1$ vector of coefficient means, and $\boldsymbol{\sigma}^2$, the $5 \times 1$ vector of coefficient variances shared by all players at that position. For each position and BIP-type, we separately estimate the posterior distribution of our parameters $\boldsymbol{\beta}$, $\boldsymbol{\mu}$ and $\boldsymbol{\sigma}^2$,

$$(7) \qquad p(\boldsymbol{\beta}, \boldsymbol{\mu}, \boldsymbol{\sigma}^2 | \mathbf{S}, \mathbf{X}) \propto p(\mathbf{S} | \boldsymbol{\beta}, \mathbf{X}) \cdot p(\boldsymbol{\beta} | \boldsymbol{\mu}, \boldsymbol{\sigma}^2) \cdot p(\boldsymbol{\mu}, \boldsymbol{\sigma}^2),$$

where $\mathbf{S}$ is the collection of all outcomes $S_{ij}$ and $\mathbf{X}$ is a collection of all location and velocity covariates $\mathbf{X}_{ij}$. We estimate the posterior distribution of all unknown parameters at each position and BIP-type using MCMC methods. Specifically, we employ a Gibbs sampling strategy [Geman and Geman (1984)] that builds upon standard hierarchical regression methodology [Gelman et al. (2003)] and data augmentation for probit models [Albert and Chib (1993)]. Additional details are provided in our supplementary materials [Jensen, Shirley and Wyner (2009)]. Our estimation procedure is repeated for each of the eighteen combinations of position and BIP type listed in Table 1, and for each of the 4 years from 2002–2005, for a grand total of $18 \times 4 = 72$ fitted models. In the next section we provide a detailed examination of our model fit for a particular position, BIP-type and year: flyballs fielded by centerfielders in 2005.

**3. Illustration of our model: flyballs to CF in 2005.** Of the 38,000 flyballs that were hit into fair territory in 2005, about 11,000 of them were caught by the CF. Of the 27,000 that were not caught by the CF, about 22,000 were caught by one of the other eight fielders and about 5000 were not caught by any fielder. The 22,000 flyballs caught by one of the other eight fielders are not treated as failures for the CF since it is unknown if the CF would have caught them had the other fielder not made the catch. These observations are treated as missing data with respect to modeling the fielding ability of the CF. The "CF-eligible" flyballs in 2005 are all flyballs that were either (1) caught by the CF or (2) not caught by any other fielder. There were exactly 15,767 CF-eligible flyballs in 2005. Figure 3 contains plots of the CF-eligible flyballs that were caught by the CF (left), and those that were not caught by the CF (right). In the right plot, data are sparse in the regions where the left fielder (LF) and right fielder (RF) play, as well as in the infield. Most of the flyballs hit to these locations were caught by the LF, RF or an infielder, and are therefore not included as CF-eligible flyballs. Additionally, we restrict ourselves only to flyballs that landed within 250 feet of the CF location for our model estimation, since traveling any larger distance to make a catch is unrealistic.

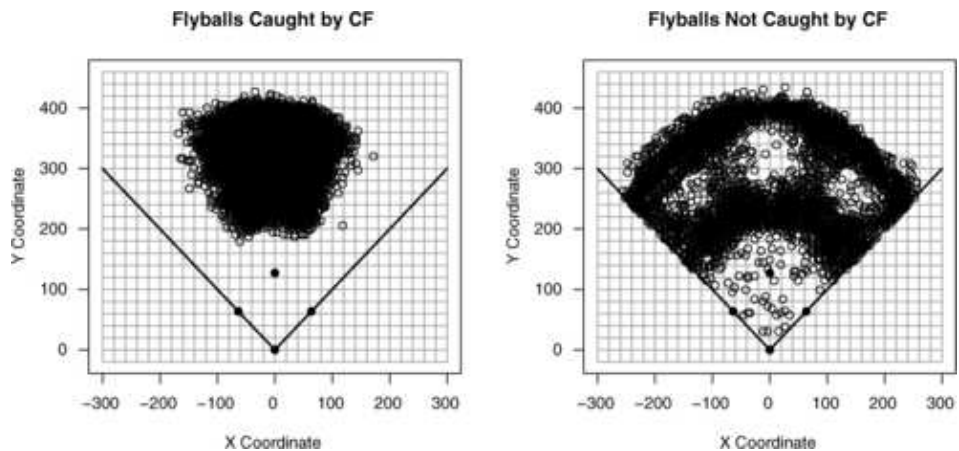

Fig. 3. *Plot of 5062 flyballs caught by center fielder (left), and 10,705 flyballs not caught by CF or any other fielder (right). Together these 15,767 points comprise the set of CF-eligible flyballs from 2005. However, only flyballs that fall within 250 feet of the CF location are used in our model fit, though this restriction only excludes a few flyballs located near home plate.*

3.1. *Data and model for illustration.* For each flyball, the data consist of the $(x, y)$-coordinates of the flyball location, the identity of the CF playing defense, and the velocity of the flyball, which is an ordinal variable with 3 levels, where 3 indicates the hardest-hit ball. In 2005 there were $N = 138$ unique CFs that played defense for at least one CF-eligible flyball. The number of flyballs per fielder, $n_i$, ranges from 1 to 531, and its distribution is skewed to the right. We denote the $(x,y)$-coordinate of the $j$th flyball hit to the $i$th CF as $(x_{ij}, y_{ij})$. Based on the overall distribution of these flyballs, we estimate the ideal starting position of a CF as the coordinate in the field with the highest catch probability across all CFs. This coordinate, which we call the CF centroid, was estimated to be $(0, 324)$, which is 324 feet into centerfield straight from home plate.

For the $j$th ball hit to the $i$th CF, we have the following covariates for our model fit: the distance from the flyball location to the CF centroid, $D_{ij} = \sqrt{(x_{ij} - 0)^2 + (y_{ij} - 324)^2}$, and the velocity of the flyball $V_{ij}$ which takes on an ordinal value from 1 to 3. As mentioned above, our model estimation only considers flyballs where $D_{ij} \leq 250$ feet. We also create an indicator variable for whether the flyball was hit to a location in front of the CF: $F_{ij} = I(y_{ij} < 324)$. $F_{ij} = 1$ corresponds to flyballs where the CF must move forward, whereas $F_{ij} = 0$ corresponds to flyballs where the CF must move backward. For the purpose of this illustration only, we consider a simplified version of our model that does not have interactions between these covariates. Specifically, we fit the following simplified model:

$$\begin{aligned}P(S_{ij} = 1) &= \Phi(\beta_{i0} + \beta_{i1} D_{ij} + \beta_{i2} V_{ij} + \beta_{i3} F_{ij}) \\ &= \Phi(\mathbf{X}_{ij} \cdot \boldsymbol{\beta}_i),\end{aligned} \tag{8}$$

where $\Phi(\cdot)$ is the cumulative distribution function for the standard normal distribution. In our full analysis, we fit the model with interactions from equation (2) in Section 2.3. For this illustration only, we also rescale the predictors $D_{ij}$, $V_{ij}$ and $F_{ij}$ to have a mean of zero and an sd of 0.5, so that the posterior estimates of $\boldsymbol{\beta}$ are on roughly the same scale, and to reduce the correlation between the intercept and the slope coefficients.

3.2. *Model implementation for illustration.* We use the Gibbs sampling approach outlined in our supplementary materials [Jensen, Shirley and Wyner (2009)] to fit our simplified model (8) for CF flyballs in 2005. Figure 4 displays posterior means and 95% posterior intervals for the four elements of the coefficient mean vector $\boldsymbol{\mu}$ shared across all CFs. As expected, the coefficients for distance and velocity are negative and, not surprisingly, distance is clearly the predictor that explains the most variation in the outcome. The coefficient for forward is positive, which means that it is easier for a CF to catch a flyball hit in front of him than behind him for the same distance and velocity. The intercept is positive, and is about 0.58. The intercept can be interpreted as the inverse probit probability $[\Phi(0.58) \approx 72\%]$ of catching a flyball hit to the mean distance from the CF (about 90 feet) at the mean velocity (about 2.2 on the scale 1–3).

Figure 5 displays three different estimates of $\boldsymbol{\beta}$:

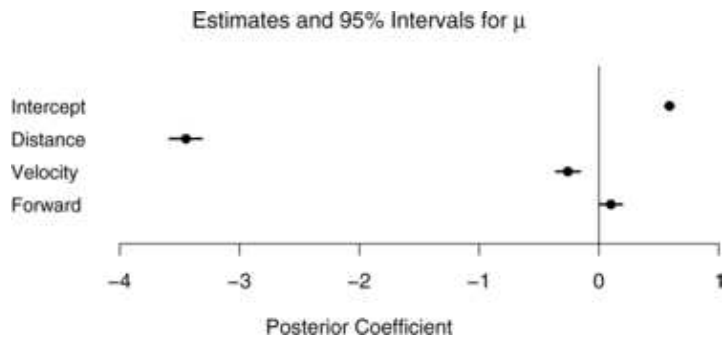

Fig. 4. *Posterior means and 95% intervals for the population-level slope coefficients $\boldsymbol{\mu}$.*

1. no pooling: $\boldsymbol{\beta}$ estimates from model with no common distribution between player coefficients,
2. complete pooling: $\boldsymbol{\beta}$ estimates from model with all players combined together for a single set of coefficients,
3. partial pooling: $\boldsymbol{\beta}$ estimates from our model described above, with separate player coefficients that share a common distribution.

From Figure 5 it is clear there that there was substantial shrinkage for the Distance and Forward coefficients, slightly less shrinkage for the Velocity coefficient, and not much shrinkage for the intercept. The posterior means for $\boldsymbol{\sigma}_k$ were 0.15, 0.28, 0.28 and 0.17 for the Intercept, Distance, Velocity and Forward coefficients, respectively. The posterior distributions of $\boldsymbol{\sigma}_k$ did not include any mass near zero, indicating that complete pooling is also not a good model, since these estimates should approach zero if there is not sufficient evidence of heterogeneity among individual players.

Figure 6 includes all $N = 138$ estimates of $\boldsymbol{\beta}_i$ $(i = 1, \ldots, N)$ with 95% intervals included. The estimates are displayed in decreasing order of $n_i$ from left to right, where the player with the most BIP observations had $n_1 = 531$ observations, and six players had just 1 observation. The players with fewer observations had their estimates shrunk much closer to the population means displayed in Figure 4, which are also drawn as horizontal lines in Figure 6, and they also had larger 95% intervals, as one would expect with fewer observations. One interesting thing to note is that a small number of players have estimated velocity coefficients that are positive, meaning they are relatively better at catching flyballs that are hit faster, and at least one player has a forward coefficient that is negative, meaning he is better at catching balls hit behind him.

To check the fit of the model graphically, we examine a number of residual plots, as shown in Figure 7. Figure 7(a) shows the histogram of the residuals,

$$r_{ij} = y_{ij} - \Phi(\mathbf{X}_{ij}\hat{\boldsymbol{\beta}}_i),$$

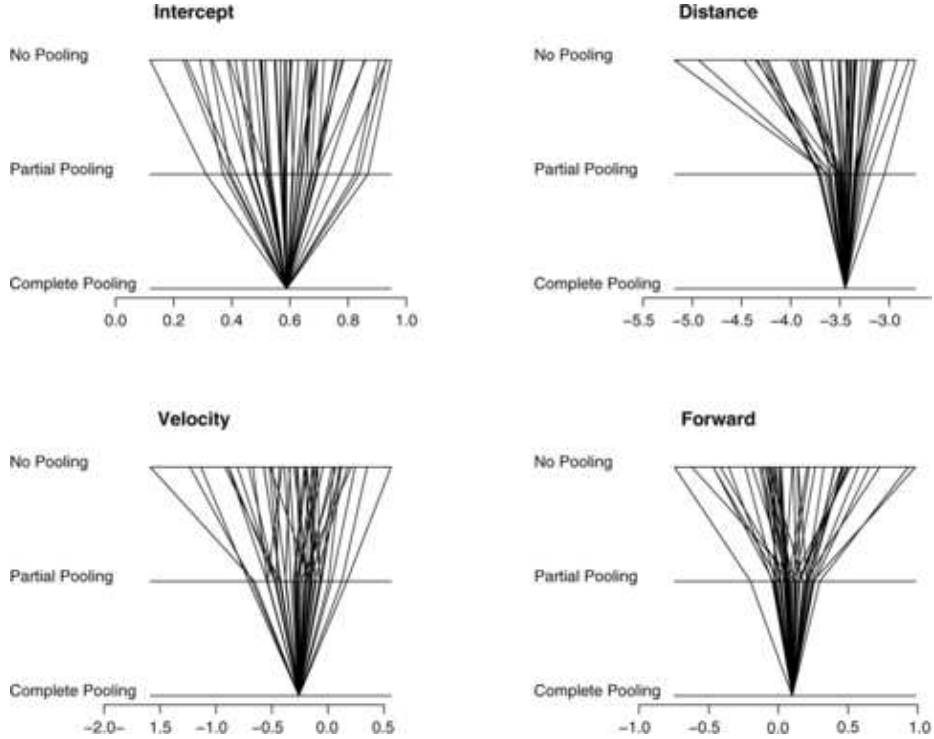

FIG. 5. *Three different estimates of $\boldsymbol{\beta}$, corresponding to no pooling, partial pooling and complete pooling. Only the 45 CFs with the largest sample sizes are included in these plots, because the no-pooling estimates for many of the CFs with little data were undefined, did not converge or were clearly unrealistic.*

for the $j$th flyball hit to the $i$th player, where $\hat{\boldsymbol{\beta}}_i$ is the posterior mean vector of the regression coefficients for player $i$. The long left tail in the Figure 7(a) histogram consists of flyballs that should have been caught (i.e., had a high predicted probability of being caught) but were not caught. Bins of residuals were constructed by ordering the residuals $r_{ij}$ in terms of the predicted probability of a catch $\Phi(\mathbf{X}_{ij}\hat{\boldsymbol{\beta}}_i)$ and then dividing the ordered residuals into equal sized bins (about 150 residuals per bin). The average of all residuals within each bin was calculated, which we call the average binned residuals. These average binned residuals are plotted as a function of predicted probabilities, which are the black points in Figure 7(b). A good model would show no obvious pattern in these average binned residuals (black dots). It appears that our model slightly overestimates the probability of catching the ball for predicted probabilities between 0% and 20%, and slightly underestimates the probability of catching the ball for predicted probabilities between 30% and 60%.

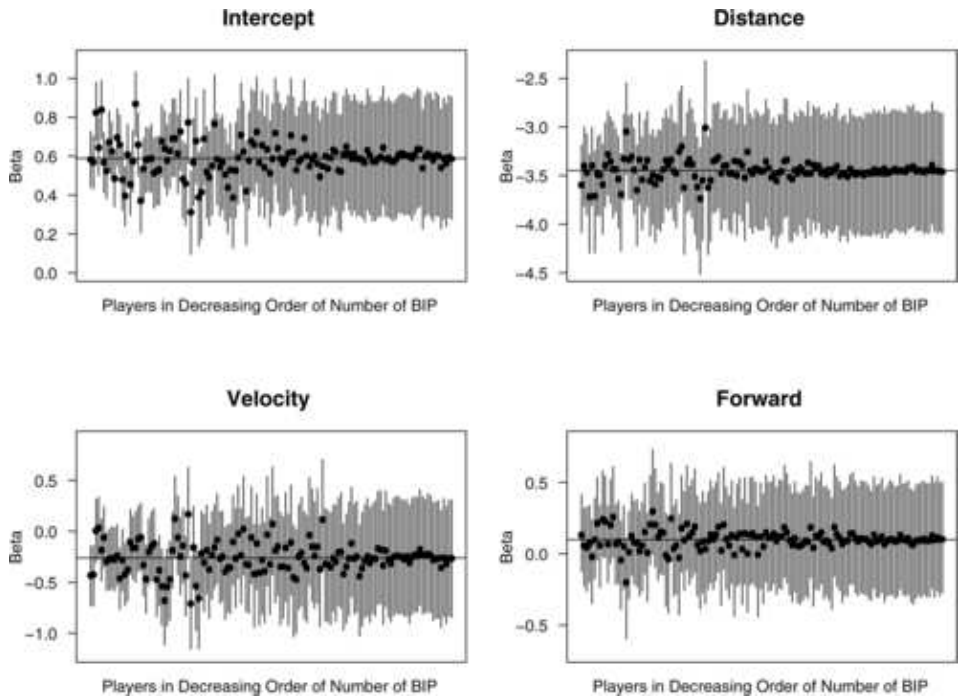

Fig. 6. *Posterior means and 95% posterior intervals for coefficients for all $N = 138$ individual players. In each plot, the distribution of $\beta_{ij}$ for each player $i$ is represented by a circle at the posterior mean and a vertical line for the 95% posterior interval. The players are displayed in decreasing order of $n_i$ from left to right, with the first player having the largest number of BIP observations ($n_1 = 531$) and the last player having the smallest number of BIP observations $n_{138} = 1$.*

In order to provide additional context to the observed residuals, we also constructed average binned residuals from 500 posterior predictive simulations of new data. These posterior predictive average binned residuals are shown as gray points in the background of Figure 7(b). We also constructed 95% posterior intervals for the average binned residuals based upon these posterior predictive simulations, and these intervals are indicated by the black lines in Figure 7(b). We see that the pattern of our observed average binned residuals is not unusual in the context of their posterior predictive distribution. In fact, we find that exactly 95 out of 100 of our observed average binned residuals fall within their 95% posterior predictive intervals, which suggests a reasonable fit. Figure 7(c) provides a different view of this same goodness-of-fit check by plotting the actual binned probabilities against the binned probabilities predicted by the model. Just as in Figure 7(b), the black points indicate the relationship from our actual data, whereas the gray points come from the same 500 posterior predictive simulations. We see that the actual binned probabilities lie approximately along the 45-degree line of equality when plotted against the predicted binned probabilities.

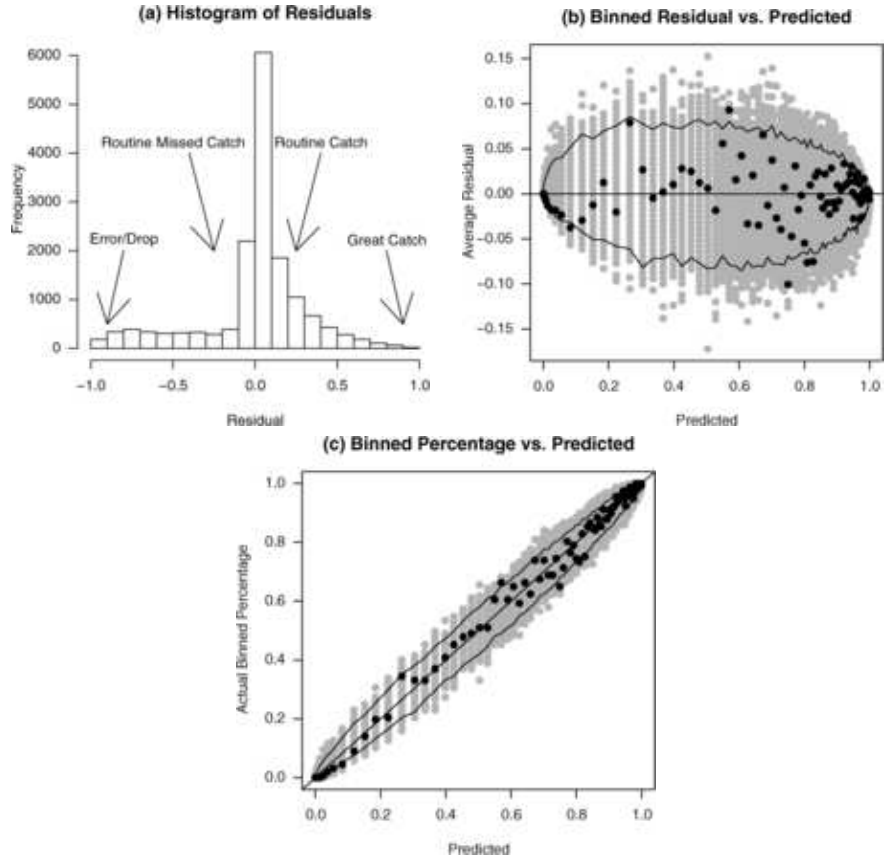

Fig. 7. *Plot* (a) *on the left shows the histogram of the fitted residuals, defined as the difference between the outcome and the expected outcome as estimated from the model using posterior means. Plot* (b) *plots average binned residuals against predicted probabilities, where the average binned residuals are the average of residuals that were binned after being ordered by the predicted probabilities. Black dots are the actual average binned residuals from our data. The gray points in the background are average binned residuals from 500 posterior predictive simulations. The black lines represent the boundaries of 95% intervals for the average binned residuals from our posterior predictive simulations. The lack of smoothness in the interval boundaries is due to randomness in our posterior predictive simulations. Plot* (c) *is constructed the same way as plot* (b), *except that the y-axis corresponds to the binned probabilities rather than binned residuals.*

We also examined the association between our residuals and individual covariates: distance, velocity, and direction, as shown in Figure 8. The plots in Figure 8 reveal no obvious patterns in the residuals with respect to the individual covariates, except possibly a slight overestimation of the proba-bility of a catch for flyballs hit at a distance of 150–200 feet from the CF. This overestimation, however, appears to be on the order of 1–2%, which is small relative to the natural variability in predictions for flyballs hit at shorter distances.

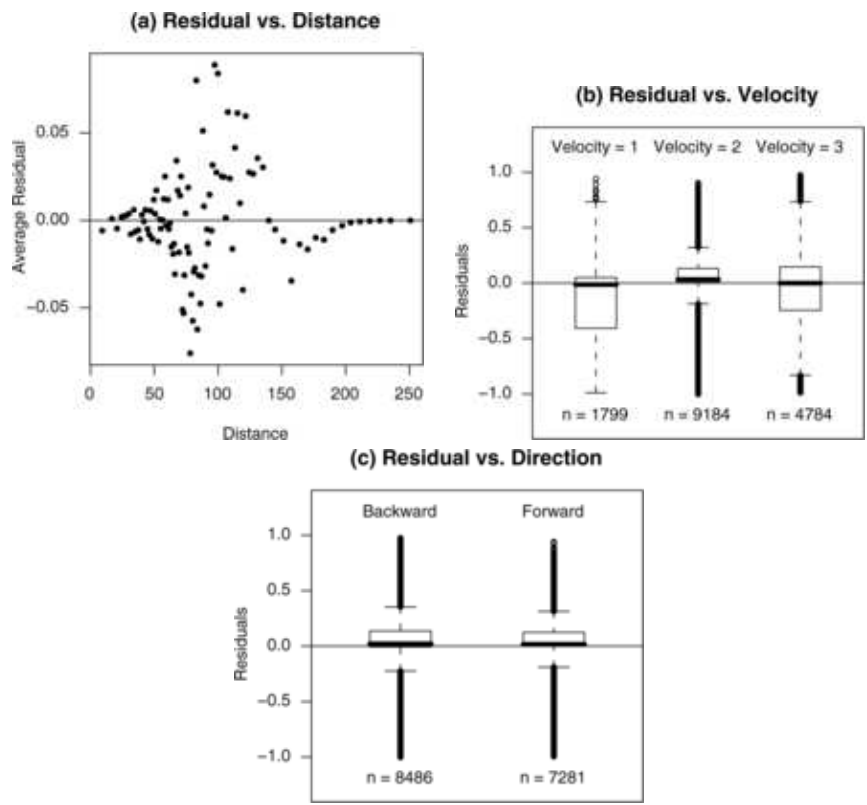

FIG. 8. *Plot* (a) *contains average binned residuals plotted vs. distance. Plot* (b) *is a boxplot of individual residuals* $r_{ij}$ *grouped by the three different levels of velocity. Plot* (c) *is a boxplot of individual residuals* $r_{ij}$ *grouped by the direction indicator: moving forwards or backward.*

We examined the shrinkage of the entire set of fitted probability curves for the whole population of CFs, shown in Figure 9. In this figure, we plot the fitted probability curves for all CFs (with fixed velocity $v = 2$ and forward $= 1$) from three different methods. Plot (a) gives the fitted probability curves estimated with no pooling—they are the curves calculated using the parameter estimates from the top horizontal line in Figure 5. Several of these curves are extreme in shape, with the most variable curves coming from players with little observed data. Plot (b) gives the curves based on parameter estimates using the probit model with our hierarchical extension presented in Section 2.5—the estimates from the "partial-pooling" middle line in Fig-ure 5. We see the stabilizing shrinkage of the partial pooling curves toward the aggregate model estimated using all data across players, which is drawn in plot (c) of Figure 9. It should be noted that the partial pooling curves are estimated using posterior means from the hierarchical model. We also explored the use of a logit model for this data, and found the model fit was similar to the probit model, which we preferred because of its computational convenience.

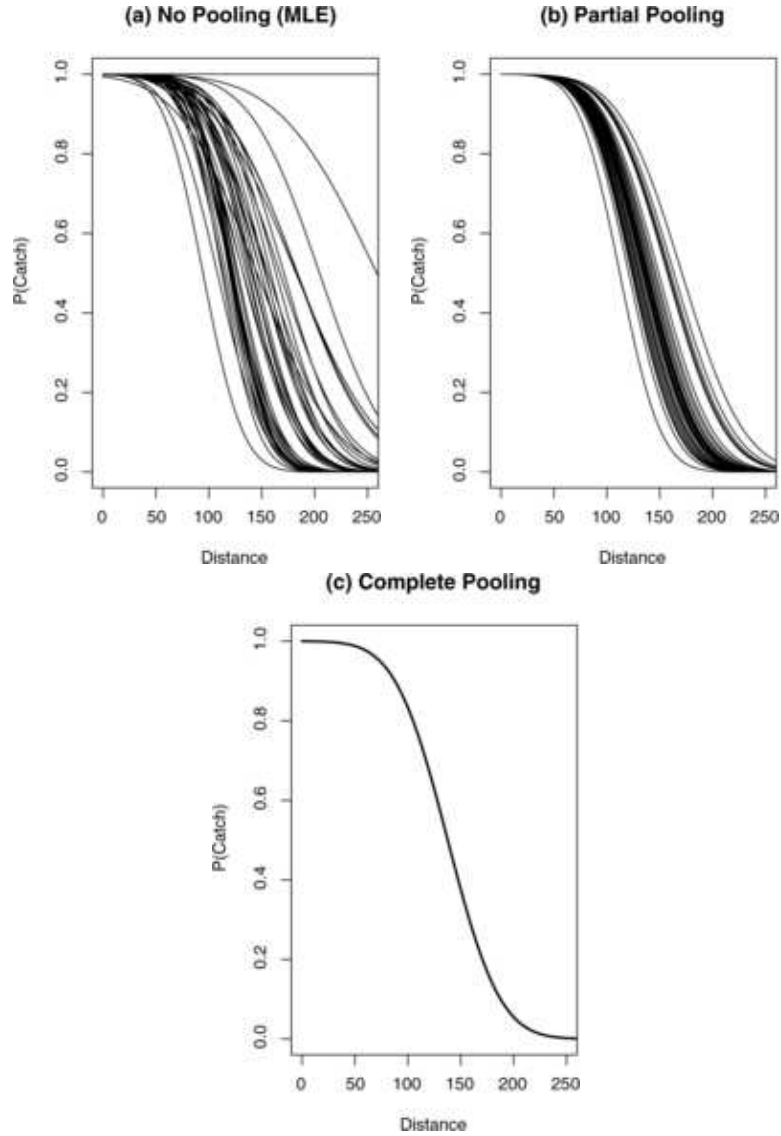

Fig. 9. *The fitted probability curves for each 2005 CF as a function of distance for flyballs hit at fixed velocity $v = 2$ in the forward direction. Plot* (a) *has curves estimated with no pooling. Plot* (b) *has the curves estimated by partial pooling via our hierarchical model (using posterior means for individual players). Plot* (c) *is the population mean curve, estimated with complete pooling.*

In addition to these overall evaluations, we also performed a range of posterior predictive checks for the fielding abilities of individual CFs. It is of interest to see if the model is accurately describing the heterogeneity between CFs, so we examined the difference in the percentage of flyballs caught between the best CF versus the worst CF. We simulated 500 posterior predictive datasets from two different models: (a) our full hierarchical model with partial pooling and (b) the complete pooling model where a single set of coefficients is fit to the data pooled across all CFs. For each of our posterior predictive datasets, we calculated the difference in the percentage of flyballs caught between the best and worst CF among the 15 CFs with the most opportunities. Figure 10 shows the density of the difference in the percentage of flyballs caught between the best and worst CF for the partial pooling model (solid density line) and the complete pooling model (dashed density line). The difference between best and worst CF from our observed data ($15.1\% = 74.1\%$ for Andruw Jones $-$ $59.0\%$ for Preston Wilson) is shown as a vertical line. We see that the actual difference from our observed data is much more likely under the partial pooling model than the complete pooling model. Not surprisingly, the complete pooling model underestimates the heterogeneity among players. Under partial pooling, however, additional variability is incorporated via the hierarchical model, so that the coefficients for each player are different, and greater differences in ability are allowed.

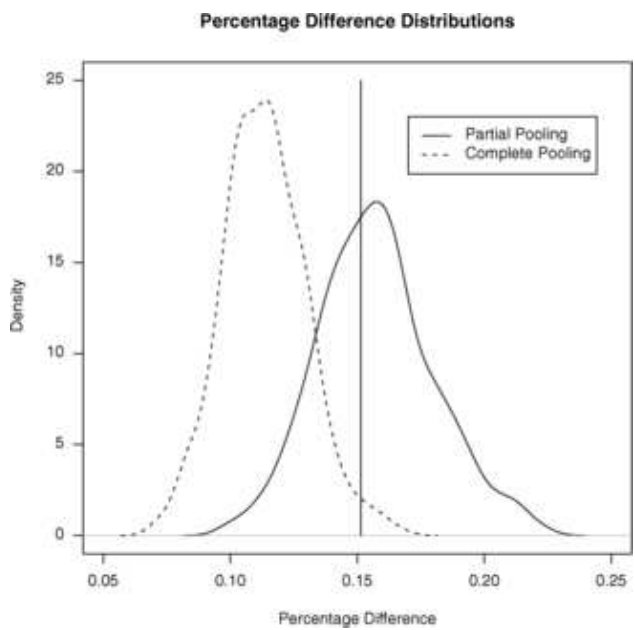

FIG. 10. *The posterior predictive density of the difference in the percentage of flyballs caught between the best CF versus the worst CF for two models. The solid-lined density represents the partial pooling model and the dotted-lined density represents the complete pooling model. These densities were estimated using 500 datasets simulated from posterior predictive distribution under these two models. The vertical line represents the difference between best and worst CF from our observed data.*

One additional concern about our model is the potential effect of outliers on the estimation of fitted probability curves. We explored the effect of a specific type of outlier: plays that were scored as fielding errors. Fielding errors are failures on BIPs that should have been fielded successfully, as judged by the official scorer for the game. Although errors contain defensive information and we prefer their inclusion in our model, the influence of these errors could be substantial since they are, by definition, unexpected results relative to the fielders' ability. We evaluated this influence on our inference for CFs by re-estimating our fitted probability models on a dataset with all fielding errors removed. These re-estimated probability curves from our Bayesian hierarchical model were essentially identical to the curves estimated with the errors included in our dataset. However, the probability curves estimated without any pooling of information were much more sensitive to the inclusion/exclusion of errors. The sharing of information between players through our hierarchical model seems to contribute additional robustness toward outlying values (in the form of errors).

**4. Converting model estimates to runs saved or cost.** In this section we use the fitted player-specific probability models from (2) and (4) for each BIP type and season to estimate the number of runs that each fielder would save or cost his team over a full season's worth of BIPs, compared to the average fielder at his position for that year.

4.1. *Comparison to aggregate curve at each position.* Our player-specific coefficients $\boldsymbol{\beta}_i$ can be used to calculate a fitted probability curve for each individual player as a function of location and velocity. For flyballs and liners, the individual fitted probability curve is denoted $p_i(x, y, v)$, the estimated probability of catching a flyball/liner hit to location $(x, y)$ at velocity $v$. For grounders, the individual fitted probability curve is denoted $p_i(\theta, v)$, the probability of successfully fielding a grounder hit at angle $\theta$ at velocity $v$. Our Gibbs sampling implementation gives us the full posterior distribution of our player-specific coefficients $\boldsymbol{\beta}_i$, which we can use to calculate the full posterior distribution of our fitted probability curves $p_i(x, y, v)$ or $p_i(\theta, v)$. Alternatively, we can calculate the posterior means $\hat{\boldsymbol{\beta}}_i$ for each $\boldsymbol{\beta}_i$ vector, and use $\hat{\boldsymbol{\beta}}_i$ to fit a single probability curve $\hat{p}_i(x,y,v)$ or $\hat{p}_i(\theta,v)$ for each player and BIP-type. For now, we focus on these single fitted probability curves, $\hat{p}_i(x,y,v)$ or $\hat{p}_i(\theta,v)$, for each player. In Section 4.2 below, we will return to an approach based on the full posterior distribution of each $\boldsymbol{\beta}_i$.

With these posterior mean fitted curves $\hat{p}_i(x,y,v)$ or $\hat{p}_i(\theta,v)$, we can quantify the difference between players by comparing their individual probabilities of making an out relative to an average player at that position. The model for the average player can be calculated in several different ways. A single probit regression model can be fit to the observed data aggregated across all players at that position to calculate the maximum likelihood estimates $\hat{\boldsymbol{\beta}}_+$, or we can use the posterior mean of the population parameters $\hat{\boldsymbol{\mu}}$. These population parameters $\hat{\boldsymbol{\beta}}_+$ can be used to calculate a fitted curve $\hat{p}_+(x,y,v)$ or $\hat{p}_+(\theta,v)$ for the average player (for flyballs/liners or grounders, respectively). Figure 11 illustrates the comparison on grounder curves between the average model for the SS position and two individual fielders.

For each possible angle $\theta$ and velocity $v$, we can calculate the difference $[\hat{p}_i(\theta,v) - \hat{p}_+(\theta,v)]$ between fielder $i$'s probability of success and the average probability of success, which is the difference in height between the individual's curve and the average curve, given in Figure 11. A positive difference at a particular angle and velocity means that the individual player is making a higher proportion of successful plays than the average fielder on balls hit to that angle at that velocity. A negative difference means that the individual player is making a lower proportion of successful plays than the average fielder on balls hit to that angle at that velocity. For our flyballs/liners models, the calculation is similar, except that we need to calculate these differences for all points around the fielder location in two dimensions. Figure 12 illustrates the comparison of probability curves between individual players and the average curve for the CF position for flyballs. For each possible location $(x, y)$ and velocity $v$, we can calculate the difference $[\hat{p}_i(x, y, v) - \hat{p}_+(x, y, v)]$ between fielder $i$'s probability of success and the average probability of success, which is the difference between the two surfaces shown in Figure 12.

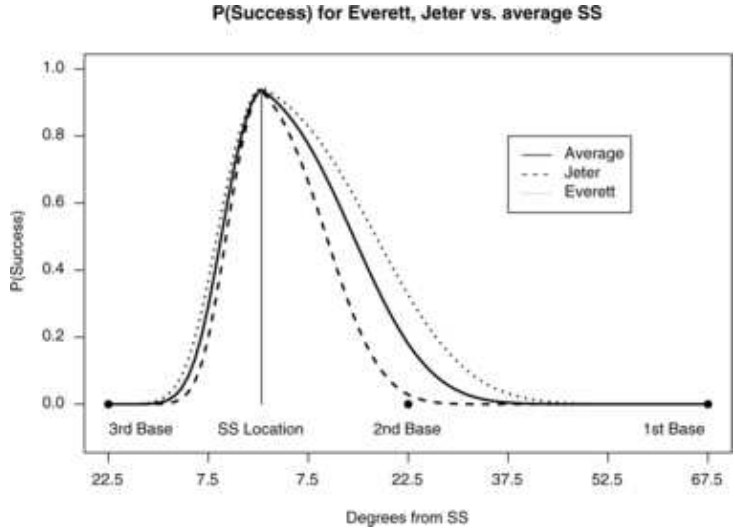

FIG. 11. *Comparison of the grounder curves of two individual SSs $\hat{p}_i(\theta,v)$ to the average SS curve $\hat{p}_+(\theta,v)$ for velocity fixed at a moderate value of $v=2$.*

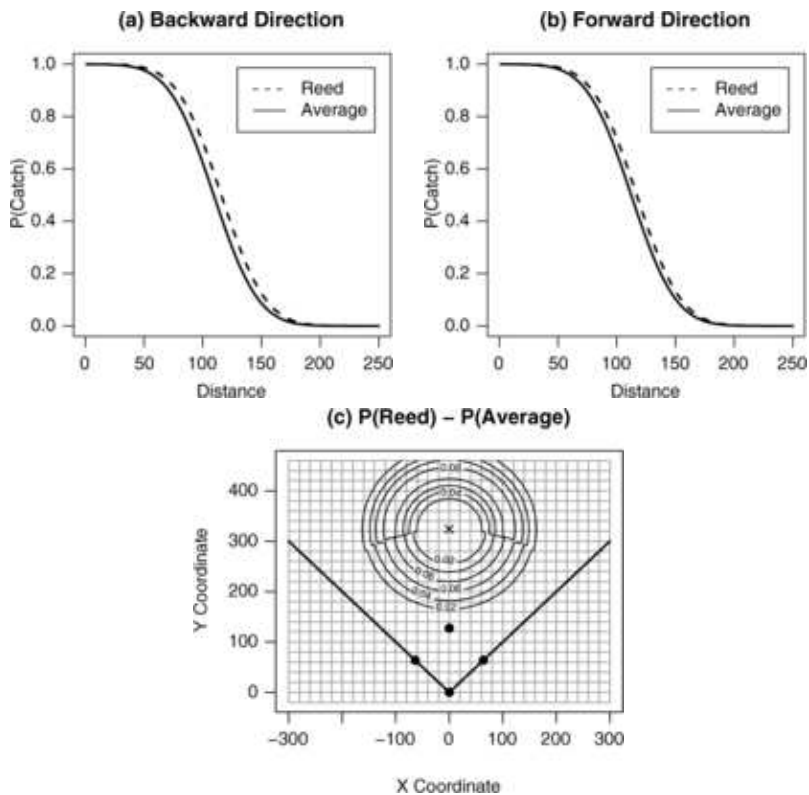

FIG. 12. *Comparison of CF curve $\hat{p}_i(x,y,v)$ for Jeremy Reed with average CF curve $\hat{p}_+(x,y,v)$ for flyballs with velocity $v = 2$ in 2005. Plot* (a) *shows the curves $\hat{p}_i(x,y,v)$ vs. $\hat{p}_+(x,y,v)$ as a function of distance moving forward from the CF location. Plot* (b) *shows the curves $\hat{p}_i(x,y,v)$ vs. $\hat{p}_+(x,y,v)$ as a function of distance moving backward from the CF location. Plot* (c) *shows a 2-dimensional contour plot of $[\hat{p}_i(x,y,v) - \hat{p}_+(x,y,v)]$. Reed's probability of catching a ball is roughly the same as the average player at short distances, but is about 8% larger at a distance of about 100 feet. Also, the difference in probability for Reed vs. the average CF is slightly larger for flyballs hit in the backward direction than for those hit in the forward direction.*

4.2. *Weighted aggregation of individual differences.* The fielding curves $\hat{p}_i(x,y,v)$ and $\hat{p}_i(\theta,v)$ for individual players give us a graphical evaluation of their relative fielding quality. For example, it is clear from Figure 11 that Adam Everett has above average range for a shortstop, whereas Derek Jeter has below average range for a shortstop. However, we are also interested in an overall numerical evaluation of each fielder which we will call "SAFE" for "Spatial Aggregate Fielding Evaluation." For flyballs or liners, one candidate value for each fielder $i$ could be to aggregate the individual differences $[\hat{p}_i(x,y,v) - \hat{p}_+(x,y,v)]$ over all coordinates $(x,y)$ and velocities $v$. For grounders, the corresponding value would be the aggregation of individual differences $[\hat{p}_i(\theta,v) - \hat{p}_+(\theta,v)]$ over all angles $\theta$ and velocities $v$. These aggregations could be carried out by numerical integration over a fine grid of values. However, these simple integrations do not take into account the fact that some coordinates $(x,y)$ or angles $\theta$ have a higher BIP frequency during the course of a season. As we saw in Figure 1, the spatial distribution of BIPs over the playing field is extremely nonuniform. Let $\hat{f}(x,y,v)$ be the kernel density estimate of the frequency with which flyballs/liners are hit to coordinate $(x,y)$, which is estimated separately for each velocity $v$. Let $\hat{f}(\theta,v)$ be the kernel density estimate of the frequency with which grounders are hit to angle $\theta$, which is estimated separately for each velocity $v$. Each fielder's overall value at a given coordinate or angle in the field should be weighted by the number of BIPs hit to that location, so that differences in ability between players in locations where BIPs are rare have little impact, and differences in ability between players in locations where BIPs are common have greater impact. Therefore, a more principled overall fielding value would be an integration weighted by these BIP frequencies,

$$\text{SAFE}_i^{fly} = \int \hat{f}(x,y,v) \cdot [\hat{p}_i(x,y,v) - \hat{p}_+(x,y,v)]\, dx\, dy\, dv,$$

$$\text{SAFE}_i^{grd} = \int \hat{f}(\theta,v) \cdot [\hat{p}_i(\theta,v) - \hat{p}_+(\theta,v)]\, d\theta\, dv.$$

As an illustration, plot (b) of Figure 13 shows the density estimate of the angle of grounders (averaged over all velocities). However, these values are still unsatisfactory because we are not addressing the fact that each coordinate or angle in the field also has a different consequence in terms of the run value of an unsuccessful play. An unsuccessful play on a pop-up to shallow left field will not result in as many runs being scored, on average, as an unsuccessful play on a fly ball to deep right field. Likewise, a grounder that goes past the first baseman down the line will result in more runs scored, on average, than a grounder that rolls past the pitcher into center field. For flyballs and liners, we estimate the run consequence of an unsuccessful play at each $(x,y)$-location in the field by first estimating two-dimensional kernel densities separately for the three different hitting events: singles, doubles and triples. We can do this using our data, in which the result of each BIP that was not fielded successfully was recorded in terms of the base that the batter reached on that BIP, which is either first, second or third base. For each $(x, y)$-coordinate in the field and velocity $v$, we use these kernel densities to calculate the relative frequency of each hitting event to each $(x, y)$-coordinate in the field with velocity $v$. We label these relative frequencies $(\hat{r}_1(x, y, v), \hat{r}_2(x, y, v), \hat{r}_3(x, y, v))$ for singles, doubles, and triples, respectively. We then calculate the run consequence for each coordinate and velocity as a function of these relative frequencies:

$$\text{(9)} \quad \hat{r}_{tot}(x, y, v) = 0.5 \cdot \hat{r}_1(x, y, v) + 0.8 \cdot \hat{r}_2(x, y, v) + 1.1 \cdot \hat{r}_3(x, y, v).$$

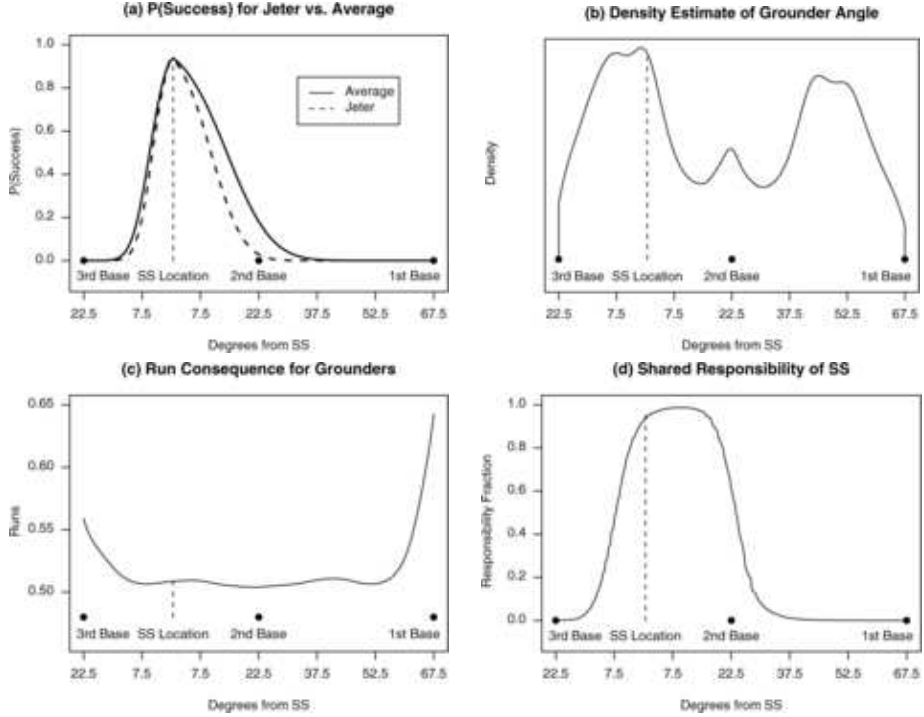

Fig. 13. *Components of our SAFE aggregation, using grounders to the SS position as an example. Plot* (a) *gives the individual grounder curve* $\hat{p}_i(\theta, v)$ *for Derek Jeter along with the the average grounder curve* $\hat{p}_+(\theta, v)$ *across all SSs for velocity fixed at a moderate value of* $v = 2$. *Plot* (b) *shows the density estimate of the BIP frequency for all grounders as a function of angle (averaged over all velocities). Plot* (c) *gives the run consequence for grounders with velocity* $v = 2$ *as a function of angle. Note the inflated consequence of grounders hit along the first and third base lines. Plot* (d) *gives the shared responsibility of the SS on grounders as a function of the angle, with a fixed velocity* $v = 2$.

The coefficients in this function come from the classical "linear weights" [Thorn and Palmer (1993)] that give the run consequence for each type of hit. These linear weights are calculated by tabulating over many seasons the average number of runs scored whenever each type of batting event occurs. From the original analysis by Palmer [Thorn and Palmer (1993)], 0.5 runs scored on average when a single was hit, 0.8 runs scored on average when a double was hit and 1.1 runs scored on average when a triple was hit. Weighting by the relative frequencies of these three events in equation (9) gives the average number of runs scored for a BIP that is not caught at every $(x,y)$-coordinate and velocity $v$. An analogous procedure produces a run consequence $\hat{r}_{tot}(\theta, v)$ for grounders at each angle $\theta$ and velocity $v$. As an example, plot (c) of Figure 13 gives the run consequence for grounders hit as a function of angle at a velocity of $v = 2$. Most grounders hit toward the middle of the field that are not fielded successfully result in singles, which have an average run value of 0.5. Only down the first and third base lines do grounders sometimes result in doubles or triples, which inflates their average run consequence. We incorporate the run consequence for each coordinate/angle as additional weights in our numerical integration,

$$\text{SAFE}_i^{fly} = \int \hat{f}(x,y,v) \cdot \hat{r}_{tot}(x,y,v) \cdot [\hat{p}_i(x,y,v) - \hat{p}_+(x,y,v)] \, dx \, dy \, dv,$$

$$\text{SAFE}_i^{grd} = \int \hat{f}(\theta,v) \cdot \hat{r}_{tot}(\theta,v) \cdot [\hat{p}_i(\theta,v) - \hat{p}_+(\theta,v)] \, d\theta \, dv.$$

In addition to run consequence, we must take into account that neighboring fielders should share the credit and blame for successful and unsuccessful plays. As an example, the difference between the abilities of two center fielders is irrelevant at a location on the field where the right fielder will always make the play. We estimate a "shared responsibility" vector for each coordinate and velocity on the field, labeled as $\hat{\mathbf{s}}(x,y,v)$ for flyballs/liners. At each coordinate $(x,y)$ and velocity $v$, we calculate the relative frequency of successful plays made by fielders at each position, and these relative frequencies are collected in the vector $\hat{\mathbf{s}}(x,y,v)$. The vector $\hat{\mathbf{s}}(x,y,v)$ has seven elements, which is the number of valid positions for flys/liners in Table 1. Similarly, we estimate a shared responsibility vector for each angle and velocity on the field, labeled as $\hat{\mathbf{s}}(\theta,v)$ for grounders. At each angle $\theta$ and velocity $v$, we calculate the relative frequency of successful plays made by fielders at each position, and these relative frequencies are collected in the vector $\hat{\mathbf{s}}(\theta,v)$. The vector $\hat{\mathbf{s}}(\theta,v)$ has four elements, which is the number of valid positions for grounders in Table 1. Plot (d) of Figure 13 gives an example of the shared responsibility of the SS position as a function of the angle, for grounders with velocity $v = 2$. The shared responsibility at each grid point for a particular player $i$ with position $pos_i$ is incorporated into their SAFE value,

$$\text{SAFE}_i^{fly} = \int \hat{f}(x,y,v) \cdot \hat{r}_{tot}(x,y,v) \cdot \hat{s}_{pos_i}(x,y,v) \tag{10}$$
$$\cdot [\hat{p}_i(x,y,v) - \hat{p}_+(x,y,v)] \, dx \, dy \, dv,$$

$$(11) \quad \text{SAFE}_i^{grd} = \int \hat{f}(\theta,v) \cdot \hat{r}_{tot}(\theta,v) \cdot \hat{s}_{pos_i}(\theta,v) \cdot [\hat{p}_i(\theta,v) - \hat{p}_+(\theta,v)] \, d\theta \, dv.$$

Figure 13 gives an illustration of the different components of our SAFE integration, using SS grounders as an example. The overall $\text{SAFE}_i$ value for a particular player $i$ is the sum of the SAFE values for each BIP type for that player's position:

$$(12) \quad \text{SAFE}_i = \text{SAFE}_i^{fly} + \text{SAFE}_i^{liner} \qquad \text{for outfielders,}$$

$$(13) \quad \text{SAFE}_i = \text{SAFE}_i^{fly} + \text{SAFE}_i^{liner} + \text{SAFE}_i^{grd} \qquad \text{for infielders.}$$

However, as noted in Section 4.1, there is no need to focus SAFE integration only on a single fitted curve $\hat{p}_+(x,y,v)$ or $\hat{p}_+(\theta,v)$ when we have the full posterior distribution of $\boldsymbol{\beta}_i$ for each player. Indeed, a more principled approach would be to calculate the integrals (10)–(11) separately for each sampled value of $\boldsymbol{\beta}_i$ from our Gibbs sampling implementation, which would give us the full posterior distribution of SAFE values for each player. In Section 5 below, we compare different individual players based upon the posterior distributions of their SAFE values.

**5. SAFE results for individual fielders.** Using the procedure described in Section 4, we calculated the full posterior distribution of $\text{SAFE}_i$ for each fielder separately for each of the 2002–2005 seasons. We will compare these posterior distributions by examining both the posterior mean and the 95% posterior interval of $\text{SAFE}_i$ for different players. The full set of year-by-year posterior means of $\text{SAFE}_i$ for each player are available for download at our project website:

http://stat.wharton.upenn.edu/~stjensen/research/safe.html.

Several fielders can have SAFE values at multiple positions in a particular year, or may have no SAFE values at all if their play was limited due to injury or retirement. In the remainder of this section we focus our attention on the best and worst individual player-years of fielding performance at each position. For each position, we focus only on players who played regularly by restricting our attention to player-years where the individual player faced more than 500 balls-in-play at that position. The following results are not sensitive to other reasonable choices for this BIP threshold.

In Table 2 we give the ten best and worst player-years at each outfield position in terms of the posterior mean of the $\text{SAFE}_i$ values. In addition to the posterior mean of $\text{SAFE}_i$, we also give the 95% posterior interval. Since each year is evaluated separately for each player, particular players can appear multiple times in Table 2. Clearly, the best fielders have positive SAFE values, indicating a positive run contribution relative to the average fielder over the course of an entire season. The worst fielders have a corresponding negative run contribution relative to the average fielder over the course of an entire season.

The magnitude of these run contributions in Table 2 are generally lower than the values obtained by previous fielding methods, such as UZR. One reason for these smaller magnitudes is the shrinkage toward the population mean imposed by our hierarchical model (Section 2.5). We also see in Table 2 that the magnitudes of the CF position are generally higher than the LF or RF positions, due to the increased number of BIPs hit toward the CF position. Another general observation from the results is the heterogeneity not only in the posterior means of $\text{SAFE}_i$ but also in the posterior variance of $\text{SAFE}_i$, as indicated by the width of the 95% posterior intervals. Indeed, even among these best/worst players (in terms of the posterior mean), we see some posterior intervals that contain zero, whereas other fielders have $\text{SAFE}_i$ intervals that are entirely above or below average.

We also examine the ten best and worst infielders at each position, where the values for corner infielders (1B and 3B) are given in Table 3 and the values for middle infielders (2B and SS) are given in Table 4. We again see a substantial difference in the magnitude of the top runs saved/cost by fielders between the different infield positions. Shortstops and second baseman have generally larger SAFE values because of the much greater number of BIPs hit to their position compared to first and third base. This increased BIP frequency to the middle infield positions seems to more than compensate for the lower run consequence of missed catches up the middle, which are almost always singles, compared to missed catches down the first or third base line, which can often be doubles or even triples. There are also substantial differences in the posterior variance of the SAFE values, as indicated by the width of the 95% posterior intervals. As with outfielders, only a subset of the best/worst infielders (in terms of the posterior mean) have posterior intervals that exclude zero, suggesting that they are significantly different than average.

One example of a player that seems to be significantly worse than average is Derek Jeter, who has some of the worst SAFE values among all shortstops. The fielding performance of Derek Jeter has always been controversial: he has been awarded several gold gloves despite being considered to have poor range by most other fielding methods. Our extremely poor SAFE value for Derek Jeter is especially interesting since our results also suggest that Alex Rodriguez has some of the best SAFE values among shortstops, especially

TABLE 2
*Outfielders in 2002–2005 with best and worst individual years of SAFE values. Posterior means and 95% posterior intervals of the SAFE values are given for each of these player-years. SAFE values can be interpreted as the runs saved or cost by that fielder's performance across an entire season*

| Ten best left fielders | | | Ten best center fielders | | | Ten best right fielders | | |
|---|---|---|---|---|---|---|---|---|
| Name and year | Post. mean | 95% post. interval | Name and year | Post. mean | 95% post. interval | Name and year | Post. mean | 95% post. interval |
| C. Crisp, 05 | 11.2 | (4.1, 17.8) | A. Jones, 05 | 11.8 | (2.2, 20.7) | J. Guillen, 05 | 6.5 | (1.8, 11.8) |
| C. Crawford, 03 | 8.5 | (1.1, 15.4) | J. Edmonds, 05 | 10.1 | (−0.5, 20.5) | R. Hidalgo, 02 | 6.4 | (−2.4, 14.1) |
| S. Stewart, 02 | 8.1 | (0.2, 16.5) | D. Erstad, 03 | 10.0 | (−1.2, 20.7) | J. D. Drew, 04 | 6.1 | (−0.5, 13.1) |
| C. Crawford, 02 | 7.7 | (−1.3, 18.6) | C. Patterson, 04 | 9.8 | (1.9, 17.9) | B. Abreu, 02 | 5.6 | (−1.6, 13.2) |
| C. Crawford, 04 | 7.6 | (1.7, 13.2) | D. Roberts, 03 | 9.6 | (1.2, 18.9) | J. Cruz, 03 | 5.5 | (−1.1, 11.2) |
| B. Wilkerson, 03 | 7.5 | (−3.2, 16.6) | A. Rowand, 02 | 9.2 | (−0.6, 20.3) | D. Mohr, 02 | 5.5 | (−3.2, 15.5) |
| P. Burrell, 02 | 6.8 | (−0.2, 14.8) | A. Jones, 03 | 9.1 | (3.2, 17.1) | S. Sosa, 04 | 5.1 | (−1.6, 14.0) |
| P. Burrell, 03 | 6.6 | (−0.9, 14.0) | M. Cameron, 03 | 8.9 | (0.3, 17.1) | A. Kearns, 02 | 4.7 | (−6.8, 16.1) |
| S. Podsednik, 05 | 6.3 | (0.4, 14.2) | A. Jones, 04 | 8.5 | (−1.2, 18.3) | J. Guillen, 03 | 4.6 | (−1.6, 11.7) |
| L. Gonzalez, 02 | 5.9 | (−3.4, 13.5) | A. Jones, 02 | 7.9 | (0.6, 15.8) | X. Nady, 03 | 4.6 | (−4.5, 13.4) |
| Ten worst left fielders | | | Ten worst center fielders | | | Ten worst right fielders | | |
| Name and year | Mean | 95% interval | Name and year | Mean | 95% interval | Name and year | Mean | 95% interval |
| M. Cabrera, 05 | −10.1 | (−18.0, −0.4) | B. Williams, 05 | −14.2 | (−23.4, −5.3) | G. Sheffield, 05 | −14.7 | (−21.6, −9.5) |
| M. Ramirez, 05 | −9.7 | (−18.4, −0.8) | B. Williams, 04 | −13.2 | (−24.5, −3.1) | V. Diaz, 05 | −6.7 | (−14.9, 2.1) |
| B. Higginson, 02 | −7.6 | (−14.0, −0.6) | K. Griffey Jr., 04 | −12.5 | (−24.4, −1.3) | B. Abreu, 05 | −6.7 | (−12.3, 0.0) |
| L. Bigbie, 03 | −6.9 | (−15.1, 1.5) | D. Roberts, 05 | −9.8 | (−21.0, 2.2) | J. Dye, 02 | −5.7 | (−14.9, 2.4) |
| R. Ibanez, 03 | −6.4 | (−12.8, 0.9) | C. Beltran, 05 | −7.5 | (−16.9, 2.8) | G. Sheffield, 04 | −5.6 | (−11.2, 0.0) |
| A. Dunn, 05 | −6.1 | (−11.2, 1.1) | J. Damon, 04 | −7.3 | (−14.4, −0.1) | B. Trammell, 02 | −5.5 | (−15.7, 7.6) |
| H. Matsui, 05 | −5.9 | (−12.4, −0.2) | C. Sullivan, 05 | −7.2 | (−20.8, 6.5) | M. Ordonez, 02 | −5.4 | (−13.0, 1.0) |
| M. Ramirez, 04 | −5.6 | (−14.8, 0.1) | B. Williams, 03 | −7.0 | (−15.5, 1.1) | J. Dye, 05 | −4.9 | (−10.1, 1.0) |
| H. Matsui, 04 | −5.5 | (−11.5, −2.0) | J. Hammonds, 02 | −6.9 | (−15.1, 1.9) | A. Huff, 03 | −4.6 | (−14.2, 6.6) |
| C. Floyd, 04 | −4.8 | (−11.1, 2.4) | G. Anderson, 04 | −6.3 | (−14.5, 3.4) | M. Cabrera, 04 | −4.0 | (−10.3, 2.6) |

BAYESIAN MODELING OF FIELDING IN BASEBALL 27

his 2003 season with the Texas Rangers. Our SAFE results seem to confirm the popular sabrmetric opinion that the New York Yankees have one of baseball's best defensive shortstops playing out of position in deference to one of the game's worst defensive shortstops. To complement these anecdotal evaluations of our results, we also compare our results to an external approach, UZR, in Section 6.

**6. Comparison to other approaches.** As mentioned in Section 1, a popular fielding measure is the Ultimate Zone Rating [Lichtman (2003)] which also evaluates fielders on the scale of run saved/cost. In general, the magnitudes of our SAFE values are generally less than UZR because of the shrinkage imposed by our hierarchical model. In fairness, it should be noted that SAFE measures the expected number of runs saves/cost, while UZR

TABLE 3

*Corner Infielders in 2002–2005 with best and worst individual years of SAFE values. Posterior means and 95% posterior intervals of the SAFE values are given for each of these player-years. SAFE values can be interpreted as the runs saved or cost by that fielder's performance across an entire season*

| Ten best 1B player-years |  |  | Ten best 3B player-years |  |  |
|---|---|---|---|---|---|
| Name and year | Mean | 95% interval | Name and year | Mean | 95% interval |
| Ken Harvey, 2003 | 5.0 | (1.5, 8.0) | Hank Blalock, 2003 | 10.0 | (4.2, 16.5) |
| Doug Mientkiewicz, 2003 | 3.4 | (−1.2, 6.5) | Sean Burroughs, 2004 | 8.9 | (3.4, 14.2) |
| Ben Broussard, 2003 | 3.2 | (1.6, 4.9) | David Bell, 2002 | 7.4 | (1.7, 13.3) |
| Eric Karros, 2002 | 2.6 | (−3.2, 7.5) | Scott Rolen, 2004 | 7.4 | (1.9, 12.1) |
| Darin Erstad, 2005 | 2.2 | (−0.8, 4.9) | Damian Rolls, 2003 | 7.2 | (0.1, 13.6) |
| Todd Helton, 2002 | 2.2 | (−3.6, 7.2) | Craig Counsell, 2002 | 6.9 | (0.9, 12.7) |
| Mike Sweeney, 2002 | 2.0 | (−2.6, 6.1) | Placido Polanco, 2002 | 5.6 | (0.3, 12.1) |
| Mark Teixeira, 2005 | 1.7 | (−1.0, 4.9) | David Bell, 2005 | 5.6 | (−0.2, 9.3) |
| Scott Spiezio, 2003 | 1.4 | (−1.2, 4.6) | Bill Mueller, 2002 | 5.4 | (−3.4, 12.6) |
| Nick Johnson, 2005 | 1.2 | (−2.0, 4.1) | Adrian Beltre, 2002 | 5.3 | (−0.4, 11.2) |

| Ten worst 1B player-years |  |  | Ten worst 3B player-years |  |
|---|---|---|---|---|
| Name and year | Mean | 95% interval | Name and year | Mean 95% interval |
| Fred McGriff, 2002 | −6.4 | (−9.4, −2.8) | Travis Fryman, 2002 | −9.4 (−15.2, −4.4) |
| Mo Vaughn, 2002 | −5.1 | (−9.7, −0.3) | Fernando Tatis, 2002 | −8.1 (−14.2, −2.0) |
| J. T. Snow, 2002 | −4.8 | (−10.1, −0.3) | Michael Cuddyer, 2005 | −7.3 (−11.4, −2.9) |
| Ryan Klesko, 2003 | −4.4 | (−8.7, −0.3) | Eric Munson, 2003 | −7.1 (−12.4, −2.8) |
| Carlos Delgado, 2005 | −4.2 | (−7.8, −0.8) | Mike Lowell, 2003 | −6.8 (−13.6, −1.6) |
| Steve Cox, 2002 | −4.0 | (−8.3, −0.3) | Wes Helms, 2004 | −6.2 (−13.8, 3.4) |
| Carlos Delgado, 2002 | −4.0 | (−8.2, 0.1) | Tony Batista, 2002 | −6.1 (−11.1, −0.9) |
| Matt Stairs, 2005 | −3.9 | (−8.3, −0.3) | Todd Zeile, 2002 | −5.8 (−11.9, −0.7) |
| Jason Giambi, 2003 | −3.8 | (−7.4, −0.2) | Chris Truby, 2002 | −5.2 (−11.7, 1.0) |
| Jeff Conine, 2003 | −3.2 | (−6.1, 0.3) | Mike Lowell, 2002 | −4.8 (−10.1, 0.8) |

TABLE 4
*Middle Infielders in 2002–2005 with best and worst individual years of SAFE values. Posterior means and 95% posterior intervals of the SAFE values are given for each of these player-years. SAFE values can be interpreted as the runs saved or cost by that fielder's performance across an entire season*

| Ten best 2B player-years | | | Ten best SS player-years | | |
|---|---|---|---|---|---|
| **Name and year** | **Mean** | **95% interval** | **Name and year** | **Mean** | **95% interval** |
| Junior Spivey, 2005    | 14.5 | (4.7, 27.1)   | Alex Rodriguez, 2003 | 13.5 | (3.5, 24.4)   |
| Chase Utley, 2005      | 10.8 | (3.1, 17.7)   | Adam Everett, 2005   | 11.5 | (1.8, 21.7)   |
| Craig Counsell, 2005   | 10.8 | (5.3, 18.0)   | Clint Barmes, 2005   | 10.8 | (−0.6, 21.5)  |
| Orlando Hudson, 2004   | 10.8 | (4.3, 16.4)   | Rafael Furcal, 2005  | 8.8  | (−0.5, 18.6)  |
| D'Angelo Jimenez, 2002 | 10.3 | (−4.9, 21.6)  | Adam Everett, 2003   | 8.7  | (−0.2, 17.7)  |
| Brandon Phillips, 2003 | 9.2  | (−0.7, 19.2)  | David Eckstein, 2003 | 8.7  | (−4.1, 20.3)  |
| Placido Polanco, 2005  | 9.0  | (2.9, 12.8)   | Bill Hall, 2005      | 8.5  | (−4.5, 23.7)  |
| Orlando Hudson, 2005   | 9.0  | (2.3, 14.8)   | Jason Bartlett, 2005 | 8.3  | (−2.8, 20.4)  |
| Mark Ellis, 2003       | 8.9  | (−0.2, 18.5)  | Jimmy Rollins, 2005  | 7.8  | (−2.6, 16.9)  |
| Brian Roberts, 2003    | 8.3  | (−0.2, 17.3)  | Alex Rodriguez, 2002 | 7.6  | (−2.1, 16.5)  |

| Ten worst 2B player-years | | | Ten worst SS player-years | | |
|---|---|---|---|---|---|
| **Name and year** | **Mean** | **95% interval** | **Name and year** | **Mean** | **95% interval** |
| Bret Boone, 2005      | −15.4 | (−22.4, −8.1) | Derek Jeter, 2005     | −18.5 | (−29.1, −9.2) |
| Luis Rivas, 2002      | −13.8 | (−20.9, −6.4) | Michael Young, 2004   | −15.6 | (−23.6, −7.2) |
| Enrique Wilson, 2004  | −12.3 | (−18.9, −6.2) | Derek Jeter, 2003     | −15.6 | (−24.8, −6.4) |
| Roberto Alomar, 2003  | −12.1 | (−19.3, −4.6) | Jhonny Peralta, 2005  | −11.4 | (−18.6, −3.5) |
| Miguel Cairo, 2004    | −10.9 | (−17.9, −3.1) | Michael Young, 2005   | −11.4 | (−20.1, −1.9) |
| Ricky Gutierrez, 2002 | −9.1  | (−18.8, 2.3)  | Derek Jeter, 2004     | −10.3 | (−20.0, −2.1) |
| Luis Rivas, 2003      | −9.0  | (−16.0, −0.9) | Deivi Cruz, 2003      | −10.1 | (−17.7, 1.2)  |
| Bret Boone, 2002      | −9.0  | (−18.2, −1.5) | Angel Berroa, 2004    | −10.0 | (−16.3, −2.4) |
| Jose Vidro, 2004      | −8.8  | (−17.7, −2.5) | Derek Jeter, 2002     | −10.0 | (−18.2, −3.6) |
| Luis Castillo, 2002   | −8.7  | (−17.1, −0.4) | Rich Aurilia, 2002    | −8.7  | (−16.6, 2.4)  |

tabulates the actual observations. However, we can still examine the correlation between the SAFE and UZR across players, which is done in Table 5 for the 423 players for which we have both SAFE and UZR values available. Note that only the 2002–2004 seasons are given because UZR values were not available for 2005. We see substantial variation between positions in terms of the correlation between SAFE and UZR. CF is the position with a high correlation, whereas 3B seems to have generally low correlation. There is also substantial variation within each position between each year. The consistency across years (or lack thereof) can be used as additional diagnostic measure for comparing our method to UZR. The problem with our comparison of methods is the lack of a gold-standard "truth" that can be used for external validation. However, one potential validation measure would be to examine the consistency of a player's SAFE value over time compared to UZR. Under the assumption that player ability is constant over time, the high consistency of a player's value over time would be indicative that our method is capturing true signal within the noise of player performance. We can measure consistency over time of SAFE with the correlation of our SAFE measures between years, as well the corresponding correlations between years of the UZR values. In Table 6 we give the correlation between the 2002 and 2003 seasons for both SAFE and UZR values, as well as the difference between these correlations. We see that overall our SAFE method does well compared to UZR, with a slightly higher overall correlation. However, there is substantial differences in performance between the different positions. The SAFE method does very well in the outfield positions, especially in CF where the correlation for our SAFE values is almost twice as high as the UZR values. However, SAFE does not perform as well in the infield positions, especially the SS position, where SAFE has a much lower correlation compared to UZR. One exception to the poor performance among infielders is the 3B position, where our SAFE values have a substantially higher correlation than UZR. We also examined the correlation between more distant years (2002 and 2004) and, as expected, the correlations are not as high for either the SAFE or UZR measures. The general conclusion from these comparisons is that our SAFE method is competitive with the popular previous method, UZR, and out-performs UZR for several positions, especially in the outfield.

An alternate way to handle the longitudinal aspect of the data would be to model the evolution of a player's fielding ability from year to year using an additional parameter or set of parameters. This type of approach has been used previously by Glickman and Stern (1998) to model longitudinal data in professional football, and could potentially allow for the modeling of a trend in the fielding ability of a baseball player across years.

TABLE 5
*Correlation between SAFE and UZR for each fielding position*

| POS | 2002 | 2003 | 2004 |
|---|---|---|---|
| 1B | 0.401 | 0.608 | 0.100 |
| 2B | 0.284 | 0.238 | 0.422 |
| 3B | 0.257 | 0.180 | 0.351 |
| CF | 0.609 | 0.546 | 0.635 |
| LF | 0.513 | 0.608 | 0.253 |
| RF | 0.410 | 0.469 | 0.392 |
| SS | 0.460 | 0.177 | 0.146 |
| Total | 0.397 | 0.440 | 0.317 |

TABLE 6
*Between year correlation for SAFE and UZR for each fielding position*

| POS | SAFE | UZR | DIFF |
|---|---:|---:|---:|
| 1B | 0.287 | 0.390 | −0.103 |
| 2B | 0.051 | 0.111 | −0.060 |
| 3B | 0.503 | 0.376 | 0.127 |
| SS | −0.030 | 0.247 | −0.277 |
| CF | 0.525 | 0.285 | 0.240 |
| LF | 0.594 | 0.548 | 0.045 |
| RF | 0.444 | 0.468 | −0.023 |
| Total | 0.372 | 0.369 | 0.003 |

**7. Discussion.** We have presented a hierarchical Bayesian probit model for estimation of spatial probability curves for individual fielders as a function of location and velocity data. Our analysis is based on data with much higher resolution of BIP location than the large zones of methods such as UZR. Our approach is model-based, which means that each player's performance is represented by a probability function with estimated parameters. One benefit of this model-based approach is that the probability of making an out is a smooth function of location in the field, which is not true for other methods. This smoothing makes the resulting estimates of our analysis less variable, since we are essentially sharing information between all points near to a fielder. Our probit models are nested within a Bayesian hierarchical structure that allows for sharing of information between fielders at a position. We have evaluated the shrinkage of curves imposed by our hierarchical model, which is intended to give improved signal for players with low sample sizes as well as reduced sensitivity to outliers, as discussed in Section 3.

We aggregate the differences between individual player curves to produce an overall measure of fielder quality which we call SAFE: spatial aggregate fielding evaluation. Our player rankings are reasonable, and when compared to previous fielding methods, namely, UZR, our SAFE values have superior consistency across years in several positions. SAFE does perform inconsistently across seasons for several other positions, especially in the infield, which merits further investigation and modeling effort. However, we note that by looking at consistency between years as a validation measure, we are assuming that player ability is actually constant over time, which may not be the case for many players. It is also worth noting that our current analysis does not take into account differences in the geography of the playing field for different parks, which could impact our outfielder evaluations. Our SAFE numerical integrations are made over a grid of points that assume the maximal park dimensions, but individual park dimensions can be quite different, with the most dramatic example being the left-field in Fenway Park. Whether or not these differences in park dimensions have a noticeable effect on our fielding evaluation will be the subject of future research.

**Acknowledgments.** Our data from Baseball Info Solutions was made available through a generous grant from ESPN Magazine. We thank Dylan Small and Andrew Gelman for helpful comments and discussion.

## SUPPLEMENTARY MATERIAL

**Gibbs sampling implementation** (DOI: 10.1214/08-AOAS228SUPP; .pdf). We provide details of our Markov chain Monte Carlo implementation, which is based on the Gibbs sampling [Geman and Geman (1984)] and the data augmentation approach of Albert and Chib (1993).

S. T. Jensen
K. E. Shirley
A. J. Wyner
Department of Statistics
The Wharton School
University of Pennsylvania
Philadelphia, Pennsylvania 19104
USA
E-mail: stjensen@wharton.upenn.edu
kshirley@wharton.upenn.edu
ajw@wharton.upenn.edu